\newif\ifDRAFT
\DRAFTtrue

\documentclass[prd ,twocolumn ,secnumarabic,dvips
,amssymb, amsmath,nobibnotes, aps, prd,superscriptaddress]{revtex4-1}
\usepackage{graphicx}

\ifDRAFT
	\usepackage[pagewise]{lineno}
	\usepackage[printwatermark]{xwatermark}
	\usepackage{xcolor}
	\usepackage{tikz}
\fi

\expandafter\ifx\csname package@font\endcsname\relax\else
 \expandafter\expandafter
 \expandafter\usepackage
 \expandafter\expandafter
 \expandafter{\csname package@font\endcsname}%
\fi

\begin{document}

\title{Construction of KAGRA: an Underground Gravitational Wave Observatory}
\author{T.~Akutsu}
\affiliation{National Astronomical Observatory of Japan, Tokyo 181-8588, Japan}
\author{M.~Ando}
\affiliation{National Astronomical Observatory of Japan, Tokyo 181-8588, Japan}
\affiliation{Research Center for the Early Universe (RESCEU), Graduate School of Science, The University of Tokyo, Tokyo 113-0033, Japan}
\affiliation{Department of Physics, Graduate School of Science, The University of Tokyo, Tokyo 113-0033, Japan}
\author{S.~Araki}
\affiliation{High Energy Accelerator Research Organization (KEK), Ibaraki 305-0801, Japan}
\author{A.~Araya}
\affiliation{Earthquake Research Institute, The University of Tokyo, Tokyo 113-0032, Japan}
\author{T.~Arima}
\affiliation{Department of Physics, Graduate School of Science, Osaka City University, Osaka 558-8585, Japan}
\author{N.~Aritomi}
\affiliation{Department of Physics, Graduate School of Science, The University of Tokyo, Tokyo 113-0033, Japan}
\author{H.~Asada}
\affiliation{Graduate School of Science and Technology, Hirosaki University, Aomori 036-8561, Japan}
\author{Y.~Aso}
\affiliation{National Astronomical Observatory of Japan, Tokyo 181-8588, Japan}
\author{S.~Atsuta}
\affiliation{Department of Physics, Graduate School of Science and Engineering, Tokyo Institute of Technology, Tokyo 152-8551, Japan}
\author{K.~Awai}
\affiliation{Institute for Cosmic Ray Research, The University of Tokyo, Chiba 277-8582, Japan}
\affiliation{Institute for Cosmic Ray Research, The University of Tokyo, Gifu 506-1205, Japan}
% ...
\collaboration{KAGRA Collaboration}
\noaffiliation
\author{K.~Agatsuma}
\affiliation{Nikhef, Amsterdam 1098 XG, The Netherlands}
\author{Y.~Akiyama}
\affiliation{Faculty of Science and Engineering, Hosei University, Tokyo 184-8584, Japan}
\author{N.~Arai}
\affiliation{Faculty of Science and Engineering, Hosei University, Tokyo 184-8584, Japan}
\author{M.~Asano}
\affiliation{Department of Physics, Graduate School of Science, Osaka City University, Osaka 558-8585, Japan}
\author{A.~Bertolini}
\affiliation{Nikhef, Amsterdam 1098 XG, The Netherlands}
\author{M.~Fujisawa}
\affiliation{Grarduate School of Science and Engineering for Education (Science), University of Toyama, Toyama 930-8555, Japan}
\author{R.~Goetz}
\affiliation{Department of Physics, The University of Florida, FL 32611, USA}
\author{J.~Guscott}
\affiliation{Department of Physics, Faculty of Science, The University of Tokyo, Tokyo 113-0033, Japan}
\author{Y.~Hashimoto}
\affiliation{Faculty of Science and Engineering, Hosei University, Tokyo 184-8584, Japan}
\author{Y.~Hayashida}
\affiliation{Astronomical Institute, Graduate School of Science, Tohoku University, Miyagi 980-8578, Japan}
\author{E.~Hennes}
\affiliation{Nikhef, Amsterdam 1098 XG, The Netherlands}
\author{K.~Hirai}
\affiliation{Graduate School of Science and Technology, Niigata University, Niigata 950-2181, Japan}
\author{T.~Hirayama}
\affiliation{Faculty of Science and Engineering, Hosei University, Tokyo 184-8584, Japan}
\author{H.~Ishitsuka}
\affiliation{Institute for Cosmic Ray Research, The University of Tokyo, Chiba 277-8582, Japan}
\affiliation{Institute for Cosmic Ray Research, The University of Tokyo, Gifu 506-1205, Japan}
\author{J.~Kato}
\affiliation{Department of Physics, Graduate School of Science and Engineering, Tokyo Institute of Technology, Tokyo 152-8551, Japan}
\author{A.~Khalaidovski}
\affiliation{Institute for Cosmic Ray Research, The University of Tokyo, Chiba 277-8582, Japan}
\affiliation{Albert-Einstein-Institut, Max-Planck-Institut f\"ur Gravitationsphysik, Hannover D-30167, Germany}
\author{S.~Koike}
\affiliation{High Energy Accelerator Research Organization (KEK), Ibaraki 305-0801, Japan}
\author{A.~Kumeta}
\affiliation{Department of Physics, Graduate School of Science and Engineering, Tokyo Institute of Technology, Tokyo 152-8551, Japan}
\author{T.~Miener}
\affiliation{Albert-Einstein-Institut, Max-Planck-Institut f\"ur Gravitationsphysik, Hannover D-30167, Germany}
\author{M.~Morioka}
\affiliation{Astronomical Institute, Graduate School of Science, Tohoku University, Miyagi 980-8578, Japan}
\author{C.~L.~Mueller}
\affiliation{Department of Physics, The University of Florida, FL 32611, USA}
\author{T.~Narita}
\affiliation{Graduate School of Science and Technology, Niigata University, Niigata 950-2181, Japan}
\author{Y.~Oda}
\affiliation{Division of Biology and Geosciences, Graduate School of Science, Osaka City University, Osaka 558-8585, Japan}
\author{T.~Ogawa}
\affiliation{Earthquake Research Institute, The University of Tokyo, Tokyo 113-0032, Japan}
\author{T.~Sekiguchi}
\affiliation{Faculty of Science and Engineering, Hosei University, Tokyo 184-8584, Japan}
\author{H.~Tamura}
\affiliation{Grarduate School of Science and Engineering for Education (Science), University of Toyama, Toyama 930-8555, Japan}
\author{D.~B.~Tanner}
\affiliation{Department of Physics, The University of Florida, FL 32611, USA}
\author{C.~Tokoku}
\affiliation{Institute of Space and Astronautical Science, Japan Aerospace Exploration Agency, Kanagawa 252-5210, Japan}
\author{M.~Toritani}
\affiliation{Department of Physics, Graduate School of Science, Osaka City University, Osaka 558-8585, Japan}
\author{T.~Utsuki}
\affiliation{Faculty of Science and Engineering, Hosei University, Tokyo 184-8584, Japan}
\author{M.~Uyeshima}
\affiliation{Earthquake Research Institute, The University of Tokyo, Tokyo 113-0032, Japan}
\author{J.~van den Brand}
\affiliation{Nikhef, Amsterdam 1098 XG, The Netherlands}
\author{J.~van Heijningen}
\affiliation{Nikhef, Amsterdam 1098 XG, The Netherlands}
\author{S.~Yamaguchi}
\affiliation{Division of Biology and Geosciences, Graduate School of Science, Osaka City University, Osaka 558-8585, Japan}
\author{A.~Yanagida}
\affiliation{Faculty of Science and Engineering, Hosei University, Tokyo 184-8584, Japan}

\begin{abstract}
Major construction and initial-phase operation of a second-generation gravitational-wave detector KAGRA has been completed. The entire 3-km detector is installed underground in a mine in order to be isolated from background seismic vibrations on the surface. This allows us to achieve a good sensitivity at low frequencies and high stability of the detector. Bare-bones equipment for the interferometer operation has been installed and the first test run was accomplished in March and April of 2016 with a rather simple configuration. The initial configuration of KAGRA is named {\it iKAGRA}. In this paper, we summarize the construction of KAGRA, including the study of the advantages and challenges of building an underground detector and the operation of the iKAGRA interferometer together with the geophysics interferometer that has been constructed in the same tunnel.
\end{abstract}

\maketitle

\section{Introduction}

Gravitational waves are ripples of spacetime radiated from dynamic motions of massive and compact 
objects, such as black holes and neutron stars, or from spacetime fluctuations in the early universe.
As gravitational waves pass though matter almost unobstructed, their observation can tell us many 
details of the nature of their sources that cannot be obtained by electromagnetic waves or other cosmic radiation.
The existence of gravitational waves was theoretically predicted by Albert Einstein in 1916. They were
not directly detected for about 100 years until the first discovery by Advanced LIGO, 
the gravitational wave observatory in the US, in 2015\,\cite{aLIGO}. 
Recently, a European gravitational
wave antenna (Advanced Virgo\,\cite{Virgo}) has begun observing and in the next few years, a Japanese
gravitational wave antenna KAGRA\,\cite{LCGT,KAGRA} will be in operation.
The new era of gravitational wave astronomy will begin with network observations by these antennas;
we can expect new astronomical knowledge with an increasing number of detected events, better parameter 
estimation accuracy for the sources (sky localization, mass, spin, distance, and orbital parameters),
and detection of gravitational waves from completely new types of sources.
These antennas (Advanced LIGO,  Advanced Virgo, and KAGRA) are called ``second-generation'' detectors,
having been upgraded from or constructed after the ``first-generation'' gravitational wave detectors
\,\cite{ref-LIGO, ref-VIRGO, ref-GEO, ref-TAMA}.

While scientists are making intense efforts towards the development of second-generation detectors, 
design studies for future antennas with better sensitivities, called  third-generation gravitational wave detectors, have begun. 
One proposal is the European antenna called the Einstein Telescope (ET)\,\cite{ET}. 
Another proposal in the US for a third-generation antenna is called Cosmic Explorer (CE)\,\cite{CE}. 
Though both are designed based on a Michelson-type laser interferometer like the second-generation detectors,
there are some differences in the design concepts.
ET will be a triangular array of three interferometers with a 10-km baseline length, and the entire system 
will be underground. 
CE will be a conventional L-shape interferometer, built either on the Earth's surface or underground, with a baseline 10 times longer than the second-generation detectors.
One of the largest differences in these design concepts is the site selection: underground or surface. 
An underground site is advantageous for gravitational wave detectors; seismic motions are smaller than a surface
site by a few orders of magnitude in typical cases. This fact is critical for the sensitivity to low-frequency gravitational wave sources, 
and long-term stable operation of a sensitive interferometer.
On the other hand, an underground site is disadvantageous from the point of view of construction cost,
requiring excavation of long tunnels and caverns for an interferometer and the vacuum system to house it.
In addition, operating a large-scale interferometer at an underground site creates new challenges 
from several practical points of view.

KAGRA~\cite{KAGRAname}, one of the second-generation detectors, is the world's first large-scale gravitational wave antenna constructed underground. 
It is located in the Kamioka underground site at Gifu Prefecture in Japan. 
At this site,  two prototype detectors, LISM (a 20-m scale interferometer, started in 1999\,\cite{LISM}) 
and  CLIO (a 100-m scale interferometer, started in 2002\,{\cite{CLIO}\cite{Agatsuma}), were constructed and operated to show 
the advantages of an underground site.
With these achievements and experiences, KAGRA was funded in 2010 and the excavation of the tunnel was started in May 2012.
In October 2015, most of the installation activities for the initial KAGRA interferometer had been completed, and
after commissioning work, the interferometer was operated  for the first time in March 2016.
Though the interferometer configuration at that time was  simplified from the final KAGRA design,
KAGRA was operated for the first time as a full 3-km-scale interferometer connected to  the data acquisition, 
transfer, and storage system.
With that configuration, we carried out a three-week test run to check the overall performance 
as an interferometric gravitational wave antenna system.

In this paper, we describe the construction and the initial-phase operation of the world's first kilometer-scale underground gravitational-wave detector, and discuss the advantages and the challenges of going underground.
In Sec.~\ref{sec:2}, we briefly explain the history of our site search and review the construction process. In Sec.~\ref{sec:3} and Sec.~\ref{sec:4}, we list advantages and challenges of building a gravitational-wave detector under the ground, which will serve as a useful reference in the planning of next-generation gravitational-wave detectors. In Sec.~\ref{sec:5}, we show the results from operation of our two underground detectors, iKAGRA itself and a geophysics interferometer built next to it in the same tunnel. In Sec.~\ref{sec:6}, we summarize the work.

\section{Construction of KAGRA}\label{sec:2}
\begin{figure}
\begin{minipage}{8.6cm}
\includegraphics[width=8.6cm]{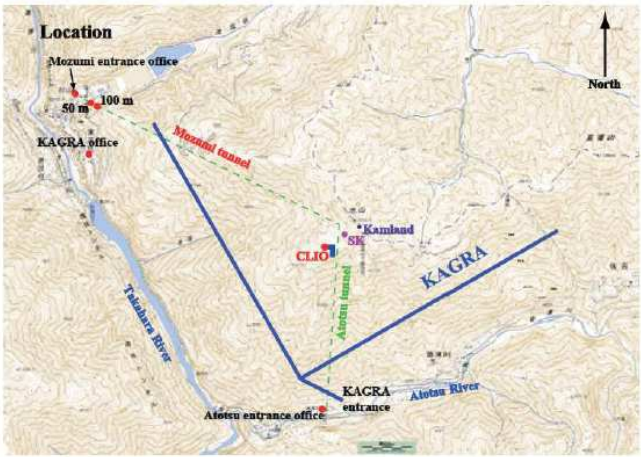}
\end{minipage} 
\caption{\label{minemap2}A map of the Kamioka mine. The blue thick lines are baselines of the KAGRA interferometer. The red points are the measurement points of the seismic motion. (See Sec.~\ref{sec:seismic}.)}
\end{figure}

The site search for KAGRA was performed in the late 90's. We consulted a company (Sumitomo Corporation, Japan) to compare the geological and environmental conditions of several candidates in Japan. One of the candidates was Mt. Tsukuba in Ibaraki Prefecture, but the cost estimate was twice as high as the Kamioka mine. Another candidate was the Kamaishi mine in Iwate Prefecture, but there was a railroad tunnel near the site. The bedrock in Kamioka, Hida gneiss, was no doubt one of the best, the estimated cost was comparatively low, and we had already started an optical experiment near the famous Super-Kamiokande that showed very low influence from seismic motion. It was a natural choice to build KAGRA in the Kamioka mine.

The excavation of the KAGRA tunnels started in 2012. Figure~\ref{minemap2} shows a map around the Kamioka mine. The geographical coordinates of the beamsplitter are 36.41 degrees North and 137.31 degrees East. The Y-arm is in the direction 28.31 degrees west of north. Though the central station and the end stations are rather close to the foot of the 1300-m high mountain, they are still at least 200\,m below the ground surface. The X and Y arms are 3-km long. Access tunnels lead to the central station and to the Y-arm end station. See Reference~\cite{UchiyamaCQG} for more details about the tunnel excavation of the KAGRA site.

A Japanese construction company, Kajima Corporation, completed the two 3-km tunnels and the central/end stations within two years (May 2012 - May 2014), making it the fastest-ever excavation work. 
Installation of the facilities proceeded in parallel with the construction work to meet timelines. This included electricity, air conditioning, water supply, anti-dust wall painting, floor treatment, crane setup, anchors, spiral steps, networks, telecommunication, laser clean room, etc. See Fig.~\ref{fig:photo1} for some photos. Coping with leaking water in the mine took time but the water issues were finally settled.
Soon after the tunnel and facility neared completion in March 2014, installation of the vacuum system and interferometer components (the vibration-isolation systems, the laser source, the optics for the interferometer, the digital control system, and so on) were started. 
\begin{figure}[t]
  \begin{center}
   \includegraphics[width=85mm]{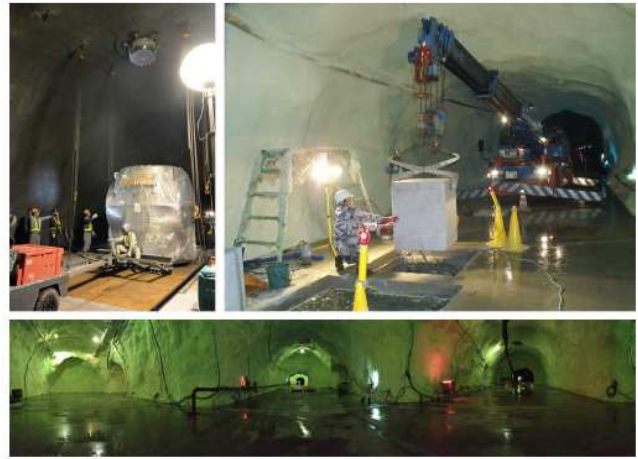}
  \end{center}
\caption{\label{fig:photo1}Some photos in the KAGRA facility. {\it Top Left}: installation of cryostat underneath the vertical borehole for the 2-story suspension tower. {\it Top Right}: installation of the granite stone for the geophysics interferometer. {\it Bottom}: panorama of the KAGRA tunnels from the central station right after paving the floor.}
\end{figure}

The KAGRA project has been split into two phases. In the first phase, called the {\it initial-phase KAGRA}, or {\it iKAGRA}, the interferometer configuration is a simple Michelson interferometer that consists of two end test masses and a beamsplitter. In the second phase, the interferometer configuration is a Michelson interferometer with cryogenic Fabry-Perot optical resonators in the two arms and also with power- and signal-recycling cavities. The detector in this stage is called the {\it baseline-design KAGRA}, or {\it bKAGRA}~\cite{KAGRA}. 

While the goal of bKAGRA is frequent observations of gravitational waves from various sources, the main goal of iKAGRA was to integrate basic subsystems and to operate the detector. It is one of the most important milestones in the KAGRA project to realize the world's first kilometer-scale underground gravitational-wave detector.

In addition to the 3-km interferometer for detecting gravitational waves, we have also built a 1.5-km interferometer in one of the KAGRA tunnels. The purpose of this interferometer is to measure Earth's strain for seismic and geodetic observations in geoscience and to feed the information back to KAGRA to characterize the behavior of the detector and to correct arm length variation for control and data analysis. It is planned to build another 1.5\,km interferometer in the other tunnel, this pair of Michelson interferometers is called a geophysics interferometer (GIF). More details about GIF are shown in Sec.~\ref{sec:GIF}.

\section{Advantages of the underground detector}\label{sec:3}
There are two main advantages to building a gravitational-wave detector in an underground site. One is its low seismic noise, which not only improves the sensitivity in the observation band but also increases the stability of the detector so that the requirements on the control system are eased. The second advantage is its low gravity gradient noise, which is one of the largest obstacles to improving the low-frequency sensitivity of next-generation detectors. In this section, we first discuss seismic noise and gravity gradient noise in the observation band. We then discuss seismic noise outside the observation band, which includes a microseismic motion at around 0.2\,Hz (caused by ocean waves) and the stability of the detector.

\subsection{Low seismic noise}\label{sec:seismic}
Prior to the excavation of the KAGRA tunnels, we investigated the seismic motion in the Kamioka mine to make a final decision of the location of the interferometer~\cite{Nishitani}. Araya {\it et al.} developed a low noise accelerometer and measured the seismic motion nearly at the center of the Kamioka mine~\cite{Araya}. The measurement revealed that the seismic motion was two orders of magnitude smaller than the typical seismic motion in a suburban area. Specifically, the power spectrum density was about $10^{-9}$m/$\sqrt{\rm Hz}$ at 1\,Hz. The measurements were repeated at the LISM site~\cite{Sato} and the CLIO site~\cite{Tomaru}, both of which were also near the center of the mine, and the results were similar. 

The locations of the KAGRA test masses need to be rather close to the foot of the mountain because the interferometer size is comparable to the mountain itself. We performed a measurement to investigate the dependence of the seismic motion as a function of the distance from the mountain surface in May 2005~{\cite{Yamamoto}\cite{Nishitani}}. The seismic motions at the various points marked red in Fig.~\ref{minemap2} including two mine entrances; Atotsu and Mozumi (south side and northwest side of the mine, respectively) were measured. The Mozumi tunnel runs straight from the Mozumi entrance to the center of the mine. We measured the seismic motion in this tunnel to investigate the position dependence of the seismic motion (50~m and 100~m from the entrance), and also measured the seismic motion at the CLIO site as a reference.

The top panel of Fig.~\ref{insideoutside} shows a typical horizontal seismic motion at each entrance of the mine. The vertical seismic motion is similar to the horizontal. Below 1~Hz, the seismic motion is comparable to that at the center of the mine (CLIO). Above 10~Hz, however, the seismic motion becomes comparable to that at Kashiwa, a suburban area northeast of Tokyo. The bottom panel of Fig.~\ref{insideoutside} shows the horizontal seismic motion in the Mozumi tunnel. The seismic motion is sufficiently small, even when the distance from the entrance is only 50~m. This means that being far from an urban area does not help reducing seismic noise in the observation band of the gravitational wave detector, but being  {\it underground} is essential for a low seismic noise level. Taking into account a safety factor, we locate all the four KAGRA test masses at least 200\,m from the surface of the mountain.

\begin{figure}
\begin{minipage}{7cm}
\includegraphics[width=7cm]{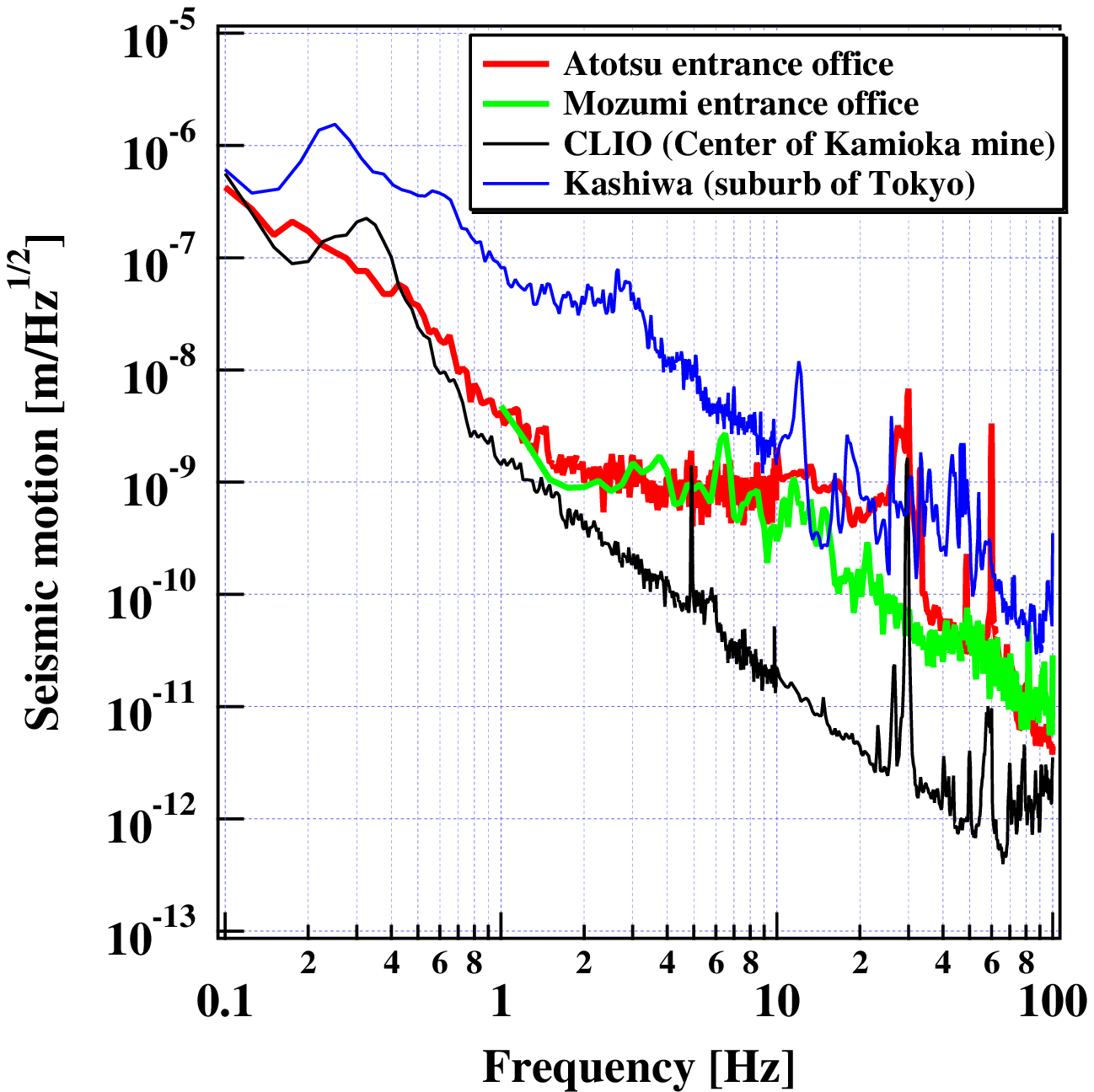}
\includegraphics[width=7cm]{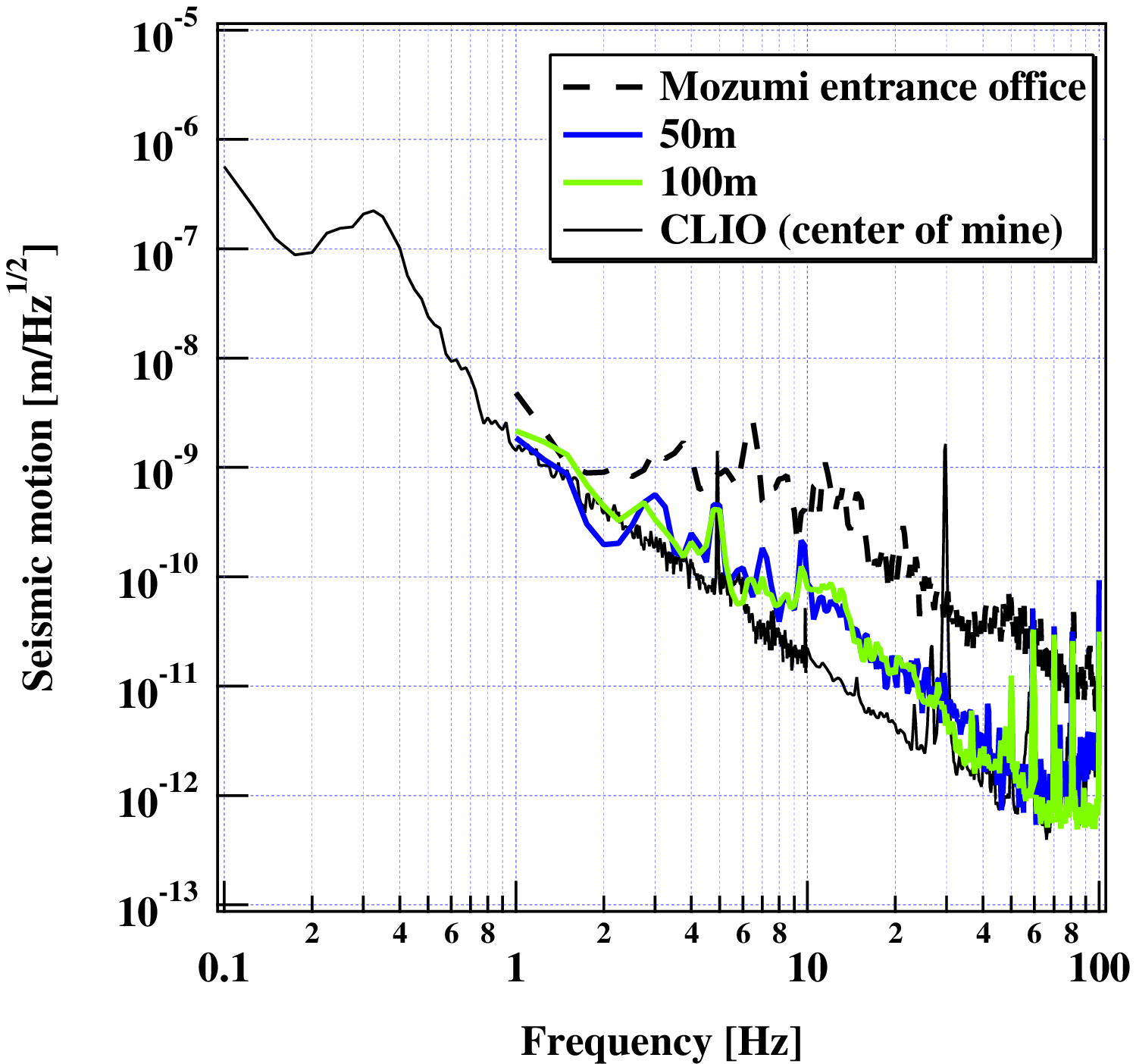}
\end{minipage} 
\caption{\label{insideoutside}{\it Top}: Typical horizontal seismic motion measured outside of Kamioka mine (Atotsu (red) and Mozumi (green) entrances). {\it Bottom}: Typical horizontal seismic motion measured in the Mozumi tunnel (the distances from the entrance are shown). The seismic motion at the CLIO site (the center of the mine, black solid) and outside of the Mozumi entrance office (black dashed) are also shown as references. It must be noted that the spectrum density at the CLIO site is mainly limited by sensor noise. In other words, the black solid lines show upper limits of seismic noise.}
\end{figure}

\subsection{Low gravity gradient noise}
\begin{figure}[htbp]
	\begin{center}
		\includegraphics[scale=0.6]{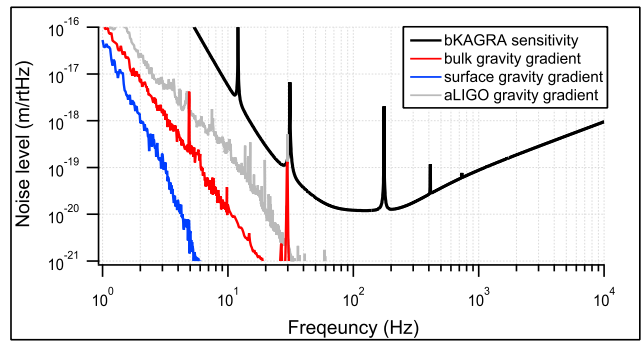}
		\caption{\label{fig:NN}Calculated gravity gradient noise in KAGRA, estimated from seismic motion measured at the CLIO site. It is smaller than that in Advanced LIGO~\cite{JanLIGO} and it does not limit the sensitivity of KAGRA in the observation band.}
	\end{center}
\end{figure}

Seismic activities change the gravitational potential around the test masses, which causes them to move. This noise is called gravity gradient noise. The source of this noise can be vibration of the ground surface, mainly due to human activity, or motion of the ground as a bulk. For a ground-based gravitational-wave detector, gravity gradient noise from the ground surface tends to be larger. Unlike the direct coupling of vibration, gravity gradient noise cannot be attenuated directly using any shield or attenuation systems. Gravity gradient noise can be a limiting noise source of ground-based gravitational wave detectors at the lower end of the observation band.
In order to reduce gravity gradient noise, it is effective to go underground since the gravity gradient noise caused by fluctuations of the ground surface is reduced \cite{Beker2011}.
Figure~\ref{fig:NN} shows the calculated gravity gradient noise spectra of the KAGRA detector. The calculation, provided by J.~Harms, is based on the seismic displacement of the CLIO site, which is very similar to that at the KAGRA site below 20 Hz, as measured by a group from NIKHEF~\cite{NIKHEFmeasurement}. The noise level is about 1 order of magnitude smaller than other sites such as LIGO and Virgo. 

There exist other kinds of gravity gradient noise in addition to that from the seismic motion. One is atmospheric gravity gradient noise~\cite{airGGN}. 
This should be decreased underground since the gravitational potential perturbation from air fluctuations  outside the building will be small. 
Another is gravity gradient noise from the water flow near the detector. This can be an issue for KAGRA as the amount of underground water in the Kamioka mine is enormous. The effect of gravity gradient noise from the water flow is currently under investigation (see App.~\ref{sec:waterGGN} for more detail).

\subsection{Stability}
While we mainly discussed seismic noise in the observation band of the detector in Sec.~\ref{sec:seismic}, the low seismic motion in Kamioka at lower frequencies also helps with stable operation of the detector. One advantage will be a high duty factor. 
Another advantage is that the loop gains of the control circuits can be kept small. In this section we discuss these two advantages and we also introduce the stability of temperature, humidity, and air pressure in the mine.

To accomplish multiple detections of gravitational waves, the duty factor of KAGRA, namely the ratio of the observation time to the entire time of the observation period, has to be sufficiently high. In KAGRA, we will have a longer startup time than other detectors because of the need to cool down and to heat up the cryogenic test masses. On the other hand, we can expect less frequent interruption of the operation since the detector is isolated from background seismic vibrations, which are responsible for most of the interruptions. The advantage of being underground is significant above $\sim1\,\mathrm{Hz}$, which will certainly help the stable operation of the interferometer. LISM, a 20-m prototype locked Fabry-Perot detector in the same mine, demonstrated 270 hours of continuous operation~\cite{LISM}. We can expect the same for KAGRA.

One of the largest technical noise sources in a gravitational-wave detector is control loop noise from auxiliary degrees of freedom (e.g. power- and signal-recycling cavity lengths). The loop noise in the observation band mainly comes from sensing noise of the controller that drives the recycling mirrors and beam splitter, and then couples to the gravitational-wave channel~\cite{loopnoise}. With the low seismic motion in the underground site, the control gain for the auxiliary degrees of freedom can be reduced. The control band of the auxiliary degrees of freedom of KAGRA is required to be lower than 20--50\,Hz in order to avoid loop noise at higher frequencies. This could be challenging but is possible in KAGRA, thanks to the low seismic motion at low frequencies in the underground site.

\begin{figure}[htbp]
	\begin{center}
		\includegraphics[scale=0.6]{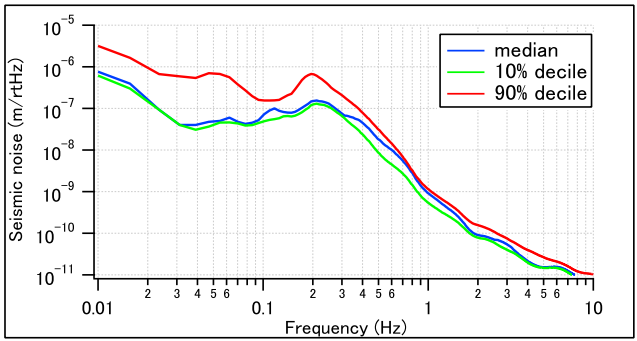}
		\caption{\label{fig:microseis}Low-frequency seismic noise in the Kamioka mine in the horizontal direction.}
		\label{microseis}
	\end{center}
\end{figure}
It should be noted, however, that the microseismic motion at around 0.2\,Hz is not low at the KAGRA site since it is not far from the Sea of Japan (about 40\,km to Toyama Bay), but is still lower than that in LIGO~\cite{LIGOseis}. Figure~\ref{microseis} shows the low-frequency seismic noise level in the horizontal direction. (The noise level in the vertical direction is almost identical.) The measurement was done by Beker et al. (NIKHEF) in 2010 using a Trillium 240 seismometer~\cite{NIKHEFmeasurement}. The same measurement was performed by Sekiguchi (ICRR) in 2014~\cite{Sekiguchi}. The blue curve shows the median of the measured seismic noise level. The red curve and green curve show the 90\,\% and 10\,\% deciles. 

CLIO performed a one-week observation run from Feb. 12--Feb. 18, 2007. Figure~\ref{fig:weatherCLIO} shows the duty factor of CLIO together with microseismic level and meteorological parameters (wind speed and air pressure outside the mine) during that week. The microseismic level was obtained from the maximum amplitude of horizontal ground velocity at the site within a 10-minute interval after a band-pass filter (0.11 - 0.5 Hz) was applied and the interval was shifted by 5 minutes. The meteorological parameters were shifted and normalized by peak-to-peak amplitudes of the period. We can see a clear correlation between the duty factor and the microseismic level while not much between the duty factor and meteorological parameters. This result implies that the underground detector is well isolated from the weather on the surface and affected mainly by ground motion originating from microseisms. The detail of the observation is shown in the CLIO elog~\cite{CLIOelog}.

\begin{figure}[htbp]
\centering
		\includegraphics[width=8.5cm]{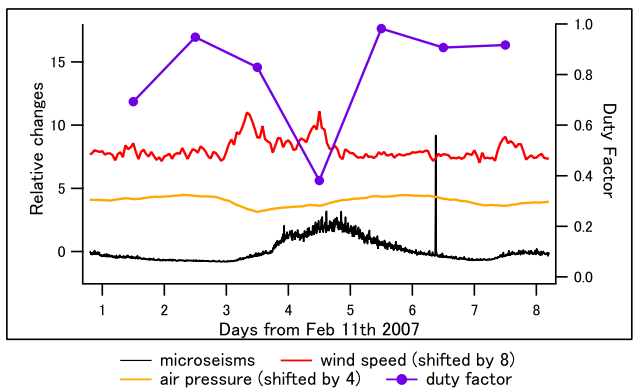}
		\caption{\label{fig:weatherCLIO}CLIO duty factor compared to microseismic level and various meteorological parameters. The meteorological parameters are obtained from the Japan Meteorological Agency website and are shifted and normalized by the peak-to-peak value in the week. The duty factor of each day is shown in the middle of the day (noon).}
\end{figure}

\begin{figure}[htbp]
\centering
		\includegraphics[width=7cm]{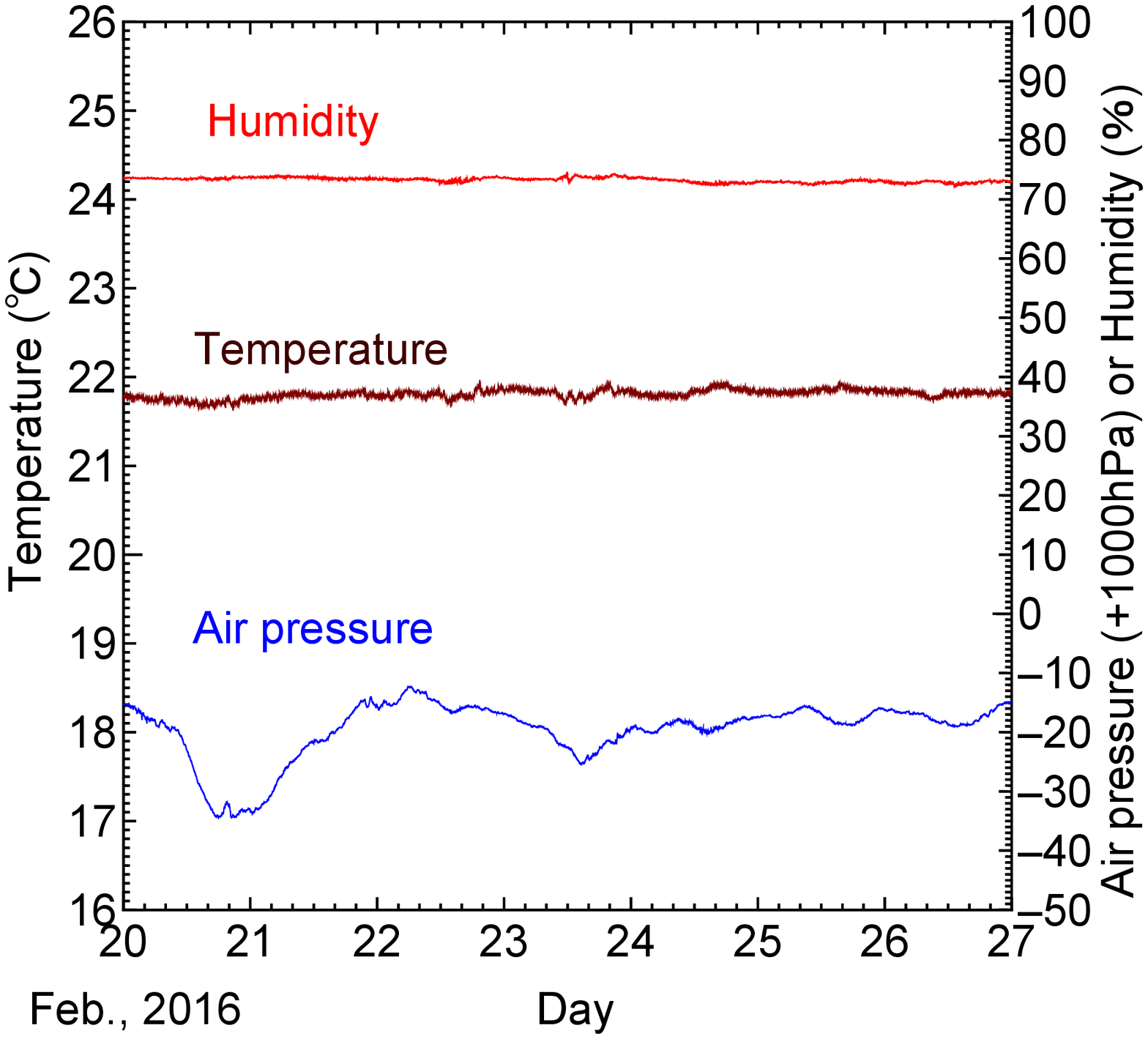}\\
\hspace{-0.7cm}
		\includegraphics[width=6.4cm]{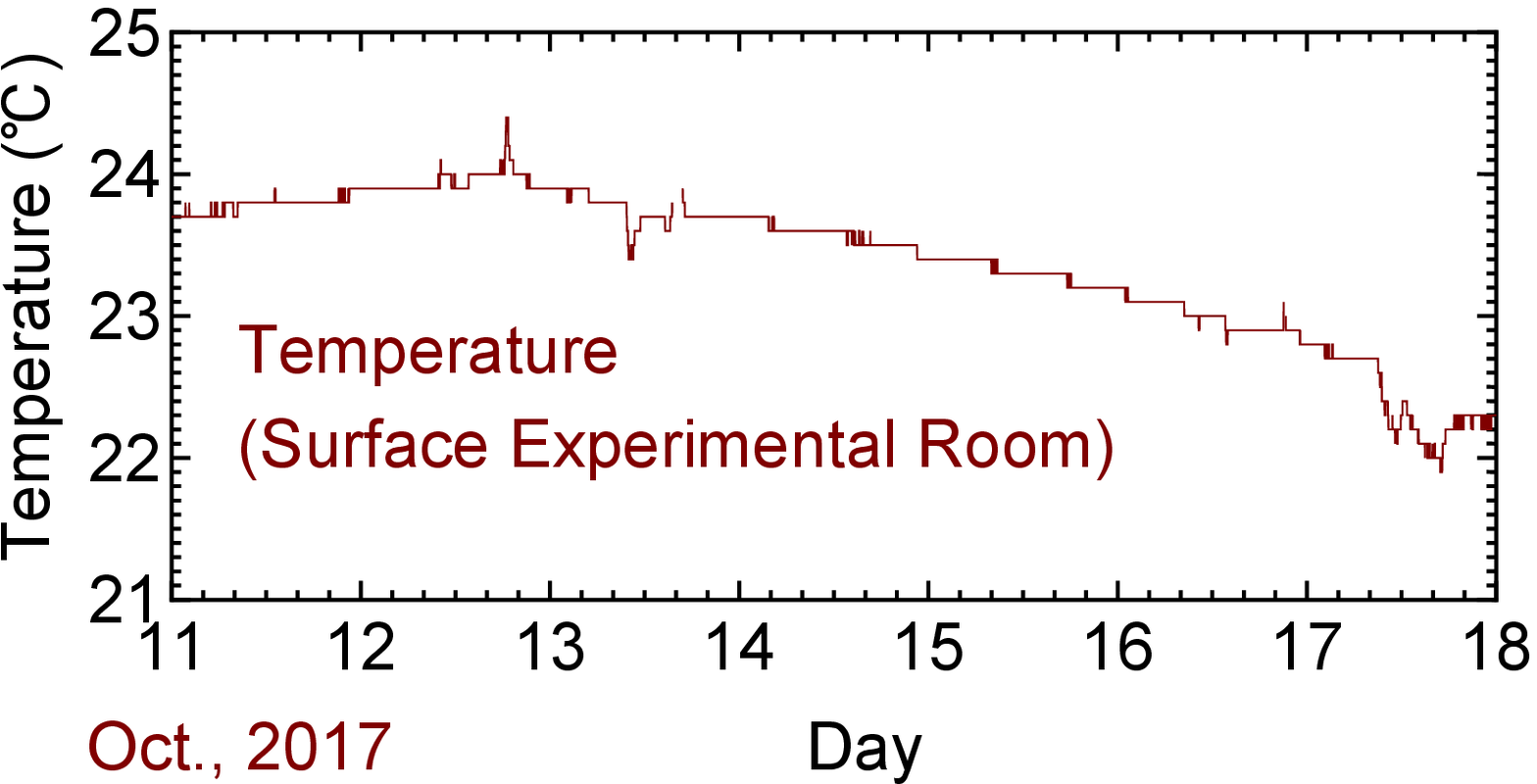}
		\caption{\label{fig:ArayaData}{\it Top}: the variation of the temperature, humidity, and air pressure in the KAGRA mine, measured in February 2016. {\it Bottom}: typical variation of the temperature in the experimental room located on surface.}
\end{figure}

The low variation of the room temperature will also help to simplify the temperature control of the whole instruments. The top panel of Fig.~\ref{fig:ArayaData} shows the variation of the temperature, humidity, and air pressure at the central area of KAGRA. The daily temperature variation over one week is as low as 0.1 degrees Celsius, while in the surface experimental room typical temperature variation (in the bottom panel) is ten times and needs to be conditioned if instruments are maintained in the constant temperature.

Small temperature variation will also help stable operation. Figure~\ref{fig:ArayaData} shows the temperature, humidity, and air pressure at the central area of KAGRA over a 1 week period. The daily temperature variation is as low as 0.1 degrees Celsius. The variation of the air pressure inside the tunnel is at the same level as the outside of the tunnel.

\section{Challenges to build the underground detector}\label{sec:4}
Offsetting the advantages we can expect by going underground, we have found various challenges in building a detector in an underground facility. In this section, we report those challenges, which we hope will be of benefit to the next-generation gravitational-wave detectors.

\subsection{GPS}
KAGRA is operated using a digital real-time control system with high-speed 
front-end computers (operating at a sampling frequency of 64\,kHz that is down-converted to 16\,kHz) and a reflective memory system that enables fast data cloning between computers that are 3\,km apart in the tunnel. In order to operate the digital system, it is important to synchronize the clocks in the computers. We use the Global Positioning System (GPS) for the synchronization but since KAGRA is inside the mine, the GPS signal does not reach the KAGRA site directly.

The GPS antenna is placed on the roof of the Hokubu-kaikan building that is 
located next to the KAGRA office outside the mine. The GPS signal is transmitted to a receiver in the Hokubu-kaikan building through an approximately 10-m electrical cable. The receiver is called the timing master and has signal receiver, splitter and transmitter functions with a delay compensation system for consistent timing across multiple items of equipment. Some of the split signals are used at the Hokubu-kaikan building to synchronize servers or other timing clients. One of the signals from the timing master is transmitted into the KAGRA mine through an approximately 7-km optical fiber cable. Inside the mine, another receiver, called the timing fanout and located in the computer room, receives the signal from the timing master. 
However, the distance of 7\,km existing between the outside and inside leads to a big time delay. The timing master and fanout establish the connection by transmitting or receiving sample signals during the first cycle of booting to determine the delay between the master and fanout. Compensation is applied between the master/fanout and the attached timing receivers. Timing receivers are implemented in the IO chassis of computers to synchronize the timing of servers and the analog-to-digital/digital-to-analog converters (ADC/DAC). The central area and both end areas are connected by a timing network with 3 timing fanouts through 3\,km optical fiber cables, each with the same compensation for the delay. Therefore all the equipment in the mine is synchronized to the GPS antenna with negligible delay.

We extract 1\,PPS signals, 65536\,Hz signals and IRIG-B signals from the GPS signal (PPS=Pulse-Per-Second, IRIG=Inter-Range-Instrumentation-Group time code). The 1\,PPS signals and 65536\,Hz signals are used to synchronize all the ADCs and DACs. The ADCs and DACs know the beginning of every 1 second interval accurately but they do not know the absolute time. The IRIG-B signal is used to synchronize all the front-end computers to absolute time. Using both 1\,PPS and IRIG-B signals tells the ADCs and DACs both the synchronization and absolute time.

\subsection{Underground water}\label{sec:water}

Water flow from the aquifer is an unavoidable issue in such a vast underground environment. 
The total amount of water flowing along the Y-arm of the KAGRA
tunnel exceeds 1,200 tons per hour in early springtime due
to melting of the heavy wintertime snow, while the amount is as little as 200\,tons per hour in summer and autumn.

It takes 2 to 7 days for rain water to start increasing the amount of underground water in the tunnel. 
The floor of the KAGRA tunnel is tilted by 1/300 in order to drain the underground water without active pumping.
The highest point is the X-end station and the lowest point is the Y-end station, so the water flows naturally from the X-end to the central station and then from the central station to the Y-end.
There are two fault crossings in the KAGRA tunnel.
One is the Ikenoyama fault crossing located in the Y-arm at around the 1800\,m point from the central station and the other is the North-20th fault crossing in the X-arm at around the 2500\,m point from the central station. (See Fig.~\ref{fig:faultcrossings}.)
While the underground water wells up throughout the tunnel, it wells up the most from these fault crossings.
In the Y-arm tunnel, we have a drainage pipe 40\,cm in diameter under the concrete floor, but this passive system alone cannot drain away all the underground water, so we have constructed a pair of additional pipes with drain pumps. 
The compulsory drainage pipes run from the 1-km point and the 2-km point toward the end of the Y-arm tunnel. 
The drain water is sent to a water reservoir located 200\,m before the Y-arm end station and then to another tunnel (owned by a local company) running 11\,m underneath the KAGRA tunnel.
This compulsory drainage system decreases the amount of water flow in the 40\,cm drainage pipe.
In the X-arm tunnel, we have increased the diameter of sections of the drainage pipe to 60\,cm so that all the underground water from the X-arm can be drained to the central station.
The drainage pipe bypasses the central station and leads to the Atotsu exit of the mine.
After all the handlings, no flood will happen in the KAGRA tunnel. Raindrops could be seen but will be handled one by one.

\begin{figure}[htbp]
	\begin{center}
		\includegraphics[scale=0.3, angle=90]{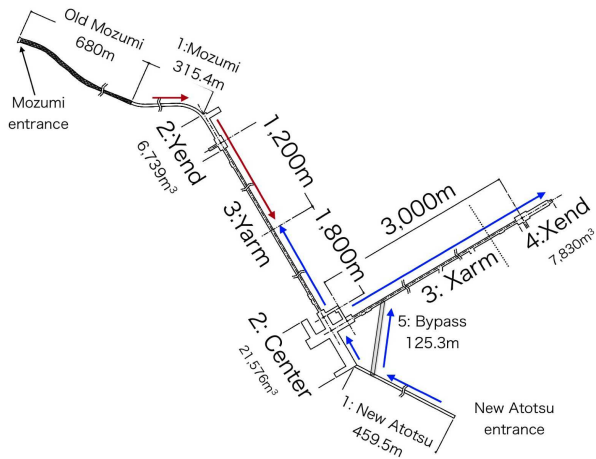}
		\caption{\label{fig:faultcrossings}KAGRA tunnel schematics. The Ikenoyama fault crossing is in the Y-arm about 1800\,m away from the center and the North-20th fault crossing is in the X-arm about 2500\,m away from the center (the red and blue arrows indicate the directions of the tunnel excavation).}
	\end{center}
\end{figure}

\subsection{Gravel layers}\label{sec:gravel}
The KAGRA tunnel was excavated by NATM (New Austrian Tunneling Method), which involves blasting the tunnel and cementing the surface wall. To make the flat tunnel floor, a layer of concrete about 200\,mm thick was cast above the gravel layer on the bedrock. Because the gravel layer tends to be deformed by heavy load, a concrete layer without gravel was placed directly on the bedrock below the KAGRA vacuum chambers and in addition the chambers were anchored to the bedrock. Even with such a structure, a long-term drift of the concrete layer causes a tilt of the chambers. 
In the iKAGRA operation, the upper limit of the drift rate of the chambers is estimated to be $<4\times10^{-6}$rad/day from the arm length data. This is manageable since we have an alignment control system of the mirrors which typically has a $\sim$mrad range. It should be noted that the drift rate will decrease in time especially if this drift is caused by the excavation of the tunnel.
GIF requires lower instrumental drift than KAGRA because long-term strain changes including a strain accumulation of a tectonic deformation need to be measured precisely. For this reason, GIF chambers are mounted on granite blocks that are directly attached to the bedrock.  Gravel was removed in part and the exposed surfaces of the bedrock were smoothed to settle the granite blocks; thicknesses of the granite blocks were 568--1100mm, depending on the depth of bedrock below the floor.

\subsection{Cleanliness and high humidity}

During the construction phase, the underground facility was pervaded with the exhaust gas from heavy machinery. The current cleanliness in the facility, however, is kept at ISO Class 6 even outside a clean booth, thanks to anti-dust paint on the wall.
As for the clean booths, KAGRA has defined two clean categories. One is an ISO class 1 category and the other is an ISO class 4 category. The former is applied only for the laser source area and the four cryostats that house the sapphire test masses. The latter is for the vacuum chambers that house silica mirrors and other optics. The system is designed to replace all of the air inside the clean booths within 2 minutes. KOACH filters manufactured by KOKEN Ltd. in Japan achieve the ISO Class 1. The KOACH filters use two kinds of filters sequentially. One, a normal HEPA filter, is a pre-filter and the other is an innovative {\it Ferina} filter that has smaller gaps made by entwining thinner fibers than in ULPA filters. Consequently, the KOACH filters are able to catch 0.1\,$\mu\mathrm{m}$ and larger-diameter particles without much constriction of the air flow. Another innovative technology in the KOACH filter is to generate parallel air flow from any position on the output surface of the filter. This flow can minimize random air movements such as vortexes that would allow particles to float for longer inside the clean area.

Another concern inside the underground facility is its high humidity. We send a total of $900\,\mathrm{m}^3/\mathrm{h}$ of dry air into the center area and the two end stations. The current relative humidity in the arm tunnel is $97\sim99\,\%$ with the temperature $17\sim18^\circ$C. During the installation of the vacuum ducts, we set up a small clean booth to house the connecting part and provided HEPA-filtered dry air using 4 heatless adsorption air driers to lower the humidity inside the booth to $70\,\%$ or less. During the assembly work on the cryostat, we provided ULPA-filtered dry air to lower the humidity inside the cryostat to $30\,\%$. (See Fig.~\ref{fig:cryohumidity}.)

\begin{figure}[htbp]
	\begin{center}
		\includegraphics[scale=0.5]{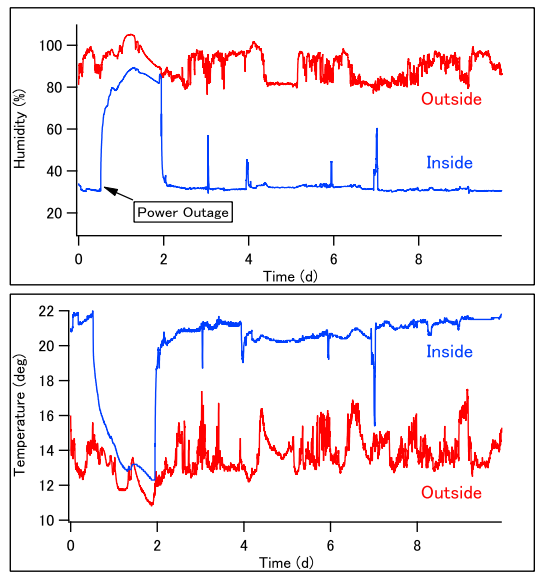}
		\caption{\label{fig:cryohumidity}Humidity inside and outside the cryostat during the assembly work. While the temperature increased, the humidity inside the cryostat was kept at 30\,\%. On the first day there was a power outage and the humidity increased for 1.5 days. Note that the humidity is different from that in Fig.~\ref{fig:ArayaData} since the temperature in the central area increased after we enclosed the facility and turned on some machines in the area.}
	\end{center}
\end{figure}

\subsection{Safety and workability}
The working environment in KAGRA has limitations due to the nature of the underground environment. The highest priority is safety. Working time is limited to the hours of 9:00--17:00, with a morning meeting held every weekday at 9:00 to discuss the schedule and to remind the on-site researchers of working precautions. For work in any area of the underground site, we mandate a buddy system, so that we can monitor the health of coworkers and help each other in case of an emergency. To monitor who is in the mine, every time we enter the tunnel, we put our nametags on a white board both in the main entrance and in the KAGRA office. (See Fig.~\ref{minemap2}.) We remove the tags when we exit the tunnel. We are planning to start using an automatic entrance monitoring system with our ID cards instead of manually putting up our nametags. Furthermore, in each underground area (the central area, X-end station and Y-end station), one safety leader is assigned. The safety leader has to carry an oxygen and carbon monoxide (CO) sensor to monitor the atmospheric conditions in the working area. In the past, during the construction work, we encountered a high-CO situation when we had to stop work. The safety leader also has to make sure no one is left in the mine alone when work is done and workers exit the tunnel.

It is mandatory to wear personal protective equipment such as helmets, work clothes with long sleeves, reflective vests and headlights (Fig.~\ref{fig:workstyle}). One of the disadvantages of the underground site is that this heavy equipment can hamper sensitive and technical tasks. Organic solvents are not allowed to be brought into the tunnel, which can also limit the efficiency of the work. When ultrasonic cleaning is required, one needs to go to University of Toyama, one of the KAGRA collaborative institutions where ultrasonic cleaning facilities are located. It takes one hour to drive from the KAGRA site to the university. In the near future a small amount of fluorinated solvents will be available in the mine, as we have recently installed a ventilation system that will exhaust solvent gas from the central area via a drafter.

\begin{figure}[htbp]
\begin{center}
\includegraphics[width=8cm]{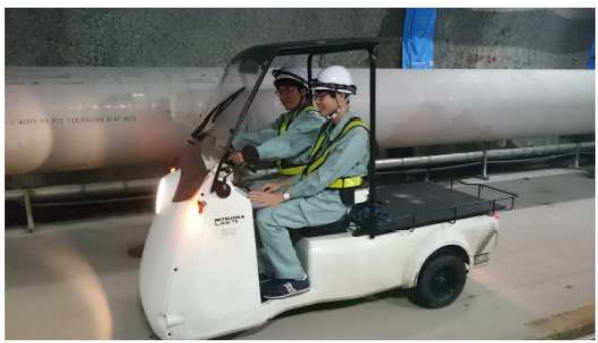}
\caption{Each worker is equipped with a helmet, working cloth, reflective vest and headlight for work inside the tunnel. Small electric carts are used for the 3-km transport from the central area to the end station.}
\label{fig:workstyle}
\end{center}
\end{figure}

The other limitation is the transportation to the mine from the KAGRA office. It takes about 10--15 minutes to drive from the KAGRA office to the tunnel entrance, via very narrow mountain roads. Only electric or diesel vehicles are allowed to enter the tunnel.  
Since a limited number of vehicles are available at once, we need to schedule the vehicles when we carry heavy loads into or out of the tunnel. 
Typically from December to early April, the roads to the mine entrance are covered by deep snow.
Snow-plows are operated by Hida city to clear away snow every morning, however, sometimes when it snows heavily, our cars parked in front of the entrance are buried in the snow, and we need to shovel out the cars to drive back. We are planning to install a snow melting system with water flow on the roads and parking area in front of the tunnel entrance.
In any case, we need to be extra cautious about transportation to and from the mine in the snow seasons.

For transportation inside the mine, we use bicycles with electric assistance and also small electric carts (Mitsuoka Motor, Like-T3, shown in Fig.~\ref{fig:workstyle}). It takes about 15 minutes to bike or drive from the central area to each end station. 

While there is an old mine tunnel connecting between the outside and the Y-end station as an emergency exit, there is no emergency exit for the X-end station. In case of an emergency, emergency rooms on the second floor of the Xend station contain oxygen, tents, temporary restrooms, water, food, and bicycles are located.  These supplies will allow five people to survive at least three days while awaiting rescue. We are considering excavating an emergency tunnel from the X-end station, or, if that is not possible, at least a ventilation pipe. The Personal Handy-phone System (PHS), fire alarm and monitoring system are available anywhere inside the tunnel, including the X- and Y-arms.

Another issue in the Kamioka area is Black bears. These bears inhabit the mountains and can attack people when they are encountered especially in autumn before they hibernate. We have set up a music speakerphone at the entrance of the KAGRA mine and remind researchers not to walk around outside a building in autumn unless needed.

\section{Operation of the underground detector}\label{sec:5}
We carried out a test run of iKAGRA to verify its overall performance. Success of this operation means that we have overcome the challenges to build the underground detector. It is the most important milestone toward the completion of bKAGRA, and to observing gravitational waves. In addition to iKAGRA, we performed a test run of GIF; though the GIF test run was performed a few months after the test operation of iKAGRA, we discuss some results of the GIF operation here as well.
\subsection{iKAGRA}\label{sec:iKAGRA}
By February 2016, we had completed the installation of the vacuum systems, clean booths, seismic isolation systems, and digital control systems to control the interferometer. In March 2016, we aligned the optics to make a 3-km Michelson interferometer, and started the test run operation. The test run was split into two periods, from March 25 to March 31 and from April 11 to April 25. Between the first and the second half of the test run, we had a little interval to improve the sensitivity and the stability of the interferometer.

\begin{figure}[htbp]
	\begin{center}
		\includegraphics[width=80mm]{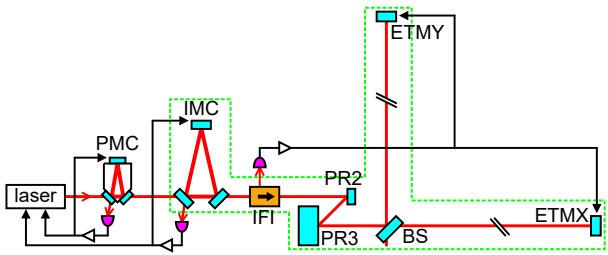}
		\caption{\label{fig:iKAGRAConfig}Interferometer configuration of iKAGRA.}
	\end{center}
\end{figure}

The interferometer configuration of iKAGRA is shown in Fig.~\ref{fig:iKAGRAConfig}. We used a single-frequency laser source with a wavelength of 1064\,nm. The laser beam was first injected into a triangular cavity called the pre-mode cleaner (PMC) and then to the suspended input mode cleaner (IMC), to suppress unwanted spatial higher order modes. The PMC was constructed from three mirrors rigidly attached to an Invar spacer. It has a round-trip length of 0.4\,m and a finesse of 197. The IMC was constructed from three mirrors suspended by a double pendulum on a vacuum-compatible vibration isolation stack table. It has a round-trip length of 53.3\,m and finesse of 540. The frequency of the laser was stabilized to the PMC length below 0.3\,mHz and the IMC length above 30\,Hz. 
The beam at IMC and the main interferometer is s-polarized. The incident power before the IMC was 270mW.

The beam transmitted by the IMC is then sent to an input faraday isolator (IFI)~\cite{IFI}, after which the beam radius is expanded by the mirrors called PR2 and PR3. PR2 and PR3 stand for power recycling 2 and 3, and will be used for the folding mirrors of the power recycling cavity in the bKAGRA phase. The main 3-km Michelson interferometer consists of the beam splitter (BS) and two end test masses (ETMX and ETMY). The designed distance between BS and ETMX is 2991.6\,m, and that between BS and ETMY is 2988.3\,m. The IFI was put on a vibration isolation stack table, and PR2 was fixed on a non-isolated table. PR3, BS, ETMX and ETMY are suspended by double pendula. Of all these suspended mirrors, PR3 was the largest (250\,mm diameter and 100\,mm thick) mirror, and was the only one with a full-sized suspension, which will also be used in the bKAGRA phase with minor modifications.

All of the optics downstream of the IMC, including the IMC, were placed inside vacuum chambers, but most of the parts were left at atmospheric pressure to allow for easier maintenance of the optics during the test run. Only the IMC and the two 3-km arm ducts are evacuated, to a pressure of $\sim10^2$\,Pa. The differential length signal of the Michelson interferometer was obtained from the reflected beam picked off by the IFI with a photo-diode in air.

The differential length signal was fed back to the ETMs via 3-km reflective memory network to lock the interferometer fringe. During the first half of the test run, we used the DC signal from the photo-diode and locked the interferometer at the mid-fringe. During the second half, we used the pre-modulation technique to lock the interferometer at the dark fringe. The dark-fringe locking resulted in a lower coupling of laser intensity noise. We also increased the bandwidth of the servo from 8\,Hz to 94\,Hz by avoiding parasitic resonances of the ETMs. In the first half of the test run, we did not have enough balancing of the mirror actuators, which resulted in the excitation of the suspension pitch modes at around 15\,Hz. 
Moreover, the servo bandwidth was automatically adjusted by monitoring the loop gain using calibration lines in the second test run. These improvements contributed to the high stability of the interferometer. The duty factor during the first half was 85.2\,\%, whereas that during the second half was 90.4\,\%. The longest lock stretch was 3.6 hours for the first half, and that for the second half was 21.3 hours. The high duty factor compared to other gravitational wave telescopes is mainly due to the simple configuration of the interferometer, but the state machine automaton {\it Guardian}~\cite{guardian_iKAGRA} also helped recovering lock quickly after lock losses.


\begin{figure}[htbp]
	\begin{center}
		\includegraphics[width=80mm]{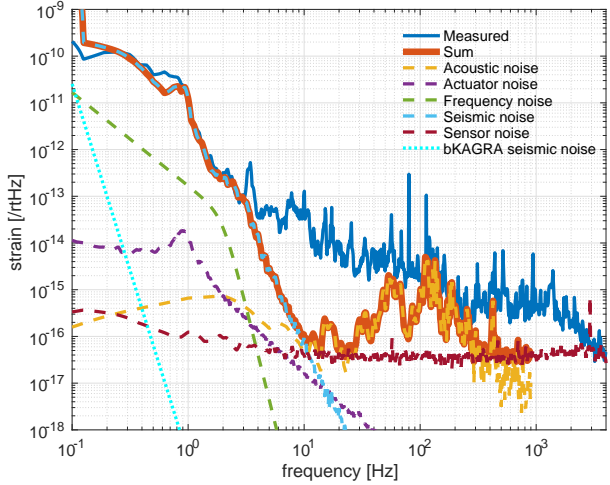}
		\caption{\label{fig:iKAGRASensitivity}The strain sensitivity of iKAGRA during the second run. An expected seismic noise level in bKAGRA is also plotted as a reference. (The noise level is about one order of magnitude smaller than other sites such as LIGO and Virgo~\cite{Rana}.) The peaks at 80\,Hz and 113\,Hz for the measured spectrum are from the calibration lines. The acoustic noise plotted here only shows the acoustic noise coupled via the BS chamber, but it is likely that the acoustic noise is the sensitivity limiting noise source also at neighboring frequencies. Actuator noise is the sum of the displacements of the mirrors from electronics noise for the actuation. Frequency noise is the estimated laser frequency noise suppressed through laser frequency stabilization using PMC and IMC. Seismic noise is the ground displacement attenuated by the mirror suspensions. Sensor noise is the sum of the ADC noise, the dark noise of the photo-diode, and the shot noise.}
	\end{center}
\end{figure}

A typical sensitivity curve for iKAGRA during the second run is shown in Fig.~\ref{fig:iKAGRASensitivity}. Below 3\,Hz, the sensitivity is limited by seismic noise. Over 100\,Hz to 3\,kHz, the sensitivity is likely to be limited by acoustic noise, mainly from the fans of the clean booths. The acoustic noise in iKAGRA was high because main mirrors were inside the vacuum chambers but at atmospheric pressure. Note that the acoustic noise curve in Fig.~\ref{fig:iKAGRASensitivity} only shows the acoustic noise coupled via the BS chamber. We turned off all the fans in the clean booths to confirm that the sensitivity of the frequency regions where the measured spectrum and the sum of all the known noise sources do not match is also limited by acoustic noise.
However, the detailed coupling mechanism still needs further investigation. At 3-5\,kHz, the sensitivity was limited by sensor noise, which mainly comes from the analog-to-digital conversion of the photo-diode output. These noise contributions will be further reduced in the bKAGRA phase by using better vibration isolation systems, a high vacuum system, and better whitening filters.

\begin{figure}[htbp]
	\begin{center}
		\includegraphics[width=85mm]{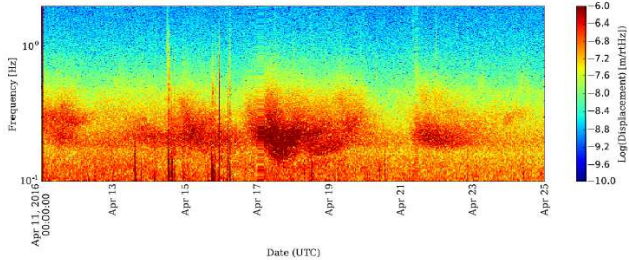}
		\caption{\label{fig:spectrogram}Spectrogram of seismic noise in the 0.1\,Hz$\sim$2\,Hz band during the second run. One can see large spikes on April 14th and 16th due to a series of large earthquakes in Kumamoto. Some data on April 17th and 21st is missing due to an issue in our data taking system.}
	\end{center}
\end{figure}

Figure~\ref{fig:spectrogram} shows a spectrogram of seismic noise in the 0.1\,Hz$\sim$2\,Hz band during the second run. In this period, a series of large earthquakes hit Kumamoto Prefecture that is about 700\,km from KAGRA. According to the United States Geological Survey, the foreshock with a magnitude 6.5 hit Kumamoto at 12:26 UTC on April 14 (2016) and the mainshock with a magnitude 7.3 hit the same area at 16:25 UTC on April 16 (2016). The peak ground acceleration (PGA) levels in the horizontal direction measured at various points in KAGRA are as follows. For the foreshock, PGAs at BS, ETMX, and ETMY were 20\,mGal, 30\,mGal, and 25\,mGal, respectively. For the mainshock, the PGA at BS was not measured as the seismometer was saturated (i.e. over 95\,mGal) as the sensor gain had been set higher than the other two. PGAs at ETMX and ETMY were 306\,mGal and 298\,mGal, respectively. After the mainshock, the suspension system of BS was in trouble and it took half a day to fix it. The earthquake shook the BS suspension and a screw to support a mirror before releasing touched one of the suspension fibers of the BS. The interferometer was not in operation during the time when several aftershock earthquakes hit Kumamoto.


During the test run, 65 people in total participated in shifts to monitor the interferometer conditions. The interferometer was monitored by at least three people on eight-hour shifts. A cumulative total of 186 people from 35 institutes contributed. The members of the shift monitored the status of the interferometer lock, mirror suspensions, data acquisition system and data transfer system to ensure that the interferometer data was properly transferred.

The data including the gravitational wave signal and environmental monitor signals were transferred to the data center at ICRR in Kashiwa and Osaka City University with latency of about 3 seconds. The amount of data was 7.5\,TB in total. We also did a hardware injection test right after the test run. The data management system and the results of the data analysis and the detector characterization will be published in separate papers.

\subsection{Geophysics interferometer}\label{sec:GIF}
\begin{figure}[htbp]
\centering
		\includegraphics[width=9cm]{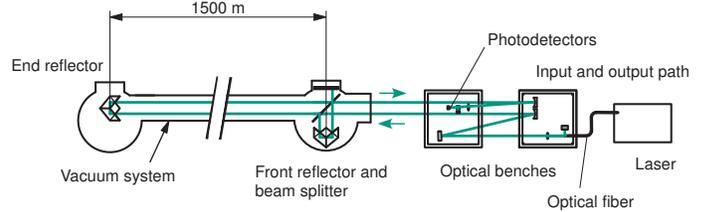}
		\caption{\label{fig:GIF}Schematic view of one of the two asymmetric Michelson interferometers of GIF.}
\end{figure}
GIF is a pair of 1.5\,km asymmetric Michelson interferometers that are to be installed in parallel with the arms of the 3-km KAGRA interferometer (Fig.~\ref{fig:GIF}). Installation of one of the GIF arms, GIF-X along the X-arm of 
KAGRA, was finished in August 2016. Two vacuum chambers, which are fixed to granite 
blocks stably installed on the bedrock as described in Sec.~\ref{sec:gravel}, house retro reflectors to 
form an interferometer. A frequency-doubled Nd:YAG laser ($\lambda$=532\,nm) is used 
as a light source. Its frequency is stabilized to an iodine-saturated absorption spectrum 
to obtain relative frequency stability $\delta\nu/\nu\simeq10^{-13}$ in Allan variance for time intervals of 10 to 1000\,s~\cite{araya1}, which corresponds 
to strain detectability $\varepsilon=\delta L/L\simeq 10^{-13}$, where $L$ is the baseline length.

The test run of GIF-X started in September 2016~\cite{GIF2017}. Fringe intensities of the quadrature interferometer 
were sampled at 50 kHz and the optical phase was calculated to obtain the variation of distance between 
two retro-reflectors separated by 1.5\,km. Typical strain variation of earth tides was clearly 
observed with GIF-X; however a slight reduction in amplitude, about 90\% of the theoretical tidal strain 
calculated by GOTIC2 including solid tides and ocean loads \cite{araya2}, was apparent. It is thought to be a 
topographic effect of the underground site judging from the similar reduction in the CLIO site \cite{araya3, araya4}. 
Strain resolution of GIF-X was estimated to be $10^{-12}- 10^{-10}$ depending on the frequency. 
Abrupt strain changes, such as might occur due to surface weather, could rarely be seen, with the exception of distant earthquakes. GIF-X mounted on the bedrock measures precise ground strain which cannot be fully predicted, for example, by standard tidal models and typical barometric response of the ground; the data are expected to reflect actual baseline of KAGRA and are necessary for the accurate characterization and correction of the arm length variation.

\section{Summary}\label{sec:6}
In this paper, we described the construction of the world's first kilometer-scale underground gravitational-wave detector, KAGRA, together with a long-baseline geophysics interferometer built in the same tunnel. We learned many lessons during the construction and we explained the advantages and challenges of going underground. We performed a first test operation of KAGRA in a simple configuration and the operation was successful. The detector will be upgraded to the cryogenic interferometer and start scientific observations in the next few years, in which the advantages of building the detector in such a quiet place under the ground will be realized.

\appendix*
\section{Possible additional issues of an underground detector}
There are also some possible additional issues in an underground site. Here we briefly discuss two of these issues.
\subsection{Gravity gradient noise of the water flow}
\label{sec:waterGGN}
While the gravity gradient noise from the seismic motion is small in the underground site, a very large amount of the underground water is flowing around the interferometer and the gravity gradient noise from the underground water can be an issue in KAGRA. As explained in Sec.~\ref{sec:water}, the underground water goes through drain pipes under the inclined tunnels, starting from the X-end station. Most of the test mass chambers are far away from the water pipes. Our largest concern is the Y-end chamber, under which we have a drainage pipe. Figure~\ref{fig:waterseismic} shows that the seismic noise level increases near the drain pipe, which was measured before we added the compulsory drainage pipe.
\begin{figure}[htbp]
	\begin{center}
		\includegraphics[width=9cm]{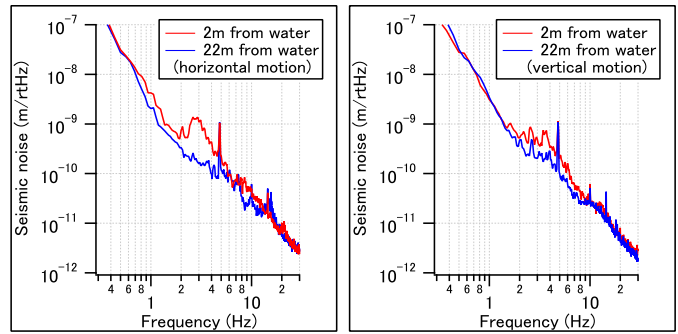}
		\caption{\label{fig:waterseismic}Seismic noise measured at two locations near the Y-end chamber; one is close to the drain pipe and the other far away from the drain pipe.}
	\end{center}
\end{figure}

Chen and Somiya made very rough models of the water flow~{\cite{YanbeiWaterGGN,SomiyaWaterGGN}}, and from these models they were not able to rule out the influence of water gravity-gradient noise on KAGRA. Further refinements of these models based on careful observations of the water flow are necessary. 

\subsection{Environmental magnetic fields}\label{sec:app2}
Environmental magnetic fields produce noise in KAGRA through the coil-magnet actuators that drive the suspended mirrors in the control system. While the magnetic noise level is known to be lower than other noise curves in the sensitivity spectrum of KAGRA, the global electromagnetic resonance in the cavity between Earth's surface and the ionosphere produces globally correlated noise components, which hamper a search of stochastic background of gravitational waves. This globally correlated magnetic noise, called Schumann resonance, is excited by lightning discharges. The Schumann resonance will be accumulated in a correlation search for the stochastic gravitational-wave background and can limit the sensitivity of global gravitational-wave detector network below $100\,\mathrm{Hz}$.

We have investigated if this kind of environmental magnetic field could be attenuated in the underground site. First we calculated the skin effect with the bedrock properties of the Kamioka mine. The electrical conductivity is $2.5\times10^{-3}\,\mathrm{\Omega}^{-1}\mathrm{m}^{-1}$ and the magnetic permeability is $1.2\times10^{-6}\,\mathrm{H/m}$. The reduction rate of the magnetic field amplitude due to the skin effect was calculated to be 0.9 at 10\,Hz and 0.7 at 100\,Hz.
We then measured the magnetic field inside and outside of the Kamioka mine on July 21-22 2016.
We arranged a set of magnetometers in the parking area in front of the KAGRA entrance and another set at the second floor of the central room inside the mine.
The magnetometers used were Metronix MFS\mbox{-}06 for the North\mbox{-}South (N\mbox{-}S) and East\mbox{-}West (E\mbox{-}W) directions and Metronix MFS\mbox{-}07e for the vertical (Z) direction.
The data were logged by an Metronix ADU\mbox{-}07e. The data were taken from on 5:00, 21 \mbox{-} July, 2016 UTC to 7:00 on the 22nd., and 7:30 \mbox{-} 8:30 22 July, 2016 UTC.
The left and right panels in Fig.~\ref{fig:mag_out} show the spectrum of the magnetic fields inside and outside the mine, respectively. The x\mbox{-}axis represents the frequency in units of Hz and the y\mbox{-}axis represents the amplitude spectrum density in units of $\mathrm{pT}/\sqrt{\mathrm{Hz}}$. The Schumann resonance can be clearly seen from its 1st mode at $7.8\,\mathrm{Hz}$ up to the 7th mode at around $50\,\mathrm{Hz}$. The coherence between the magnetic field inside and outside of the mine is known to be as high as 0.9 at the Schumann resonances~\cite{atsuta}.

\begin{figure}
\begin{center}
\includegraphics[width=0.8\linewidth]{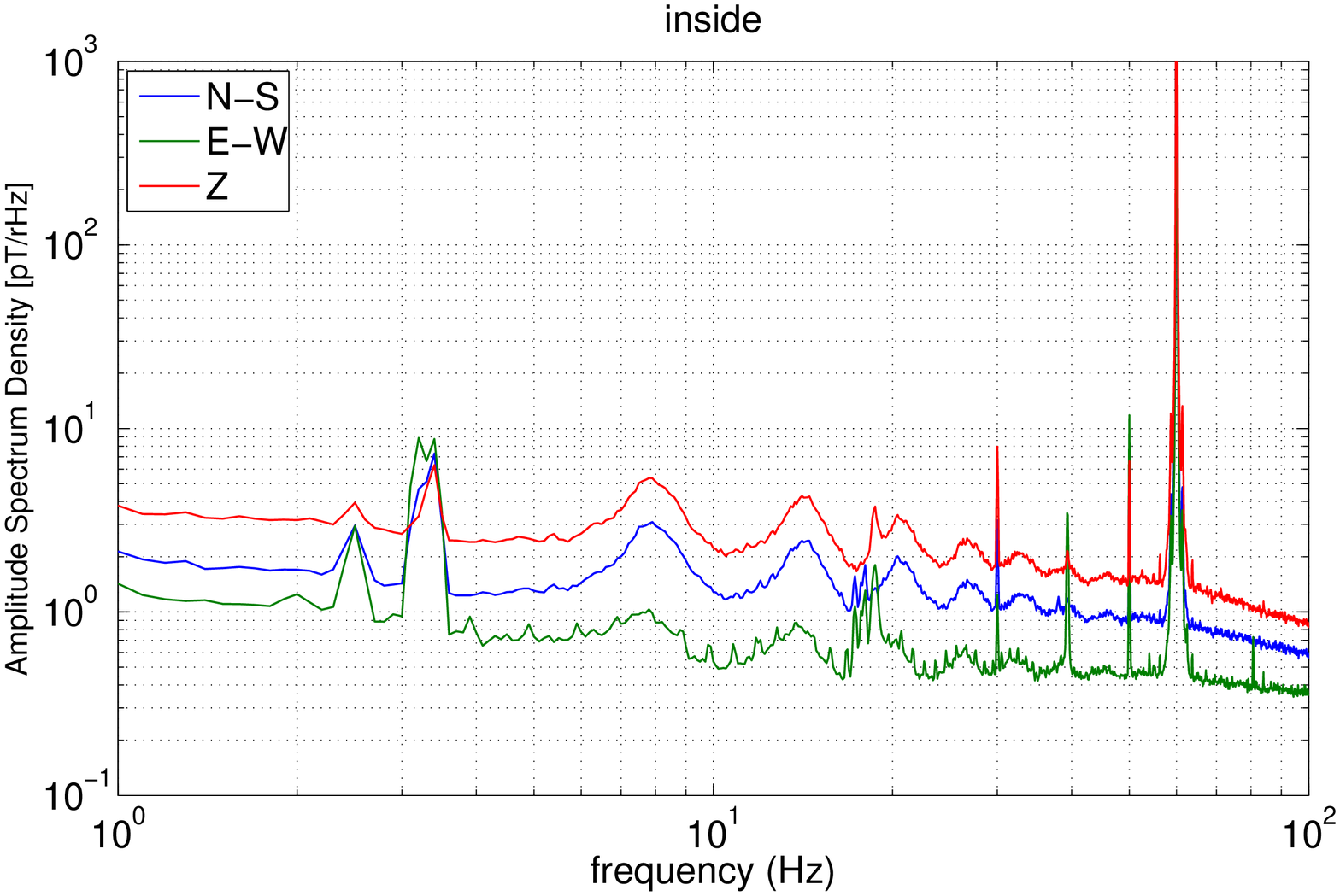}
\includegraphics[width=0.8\linewidth]{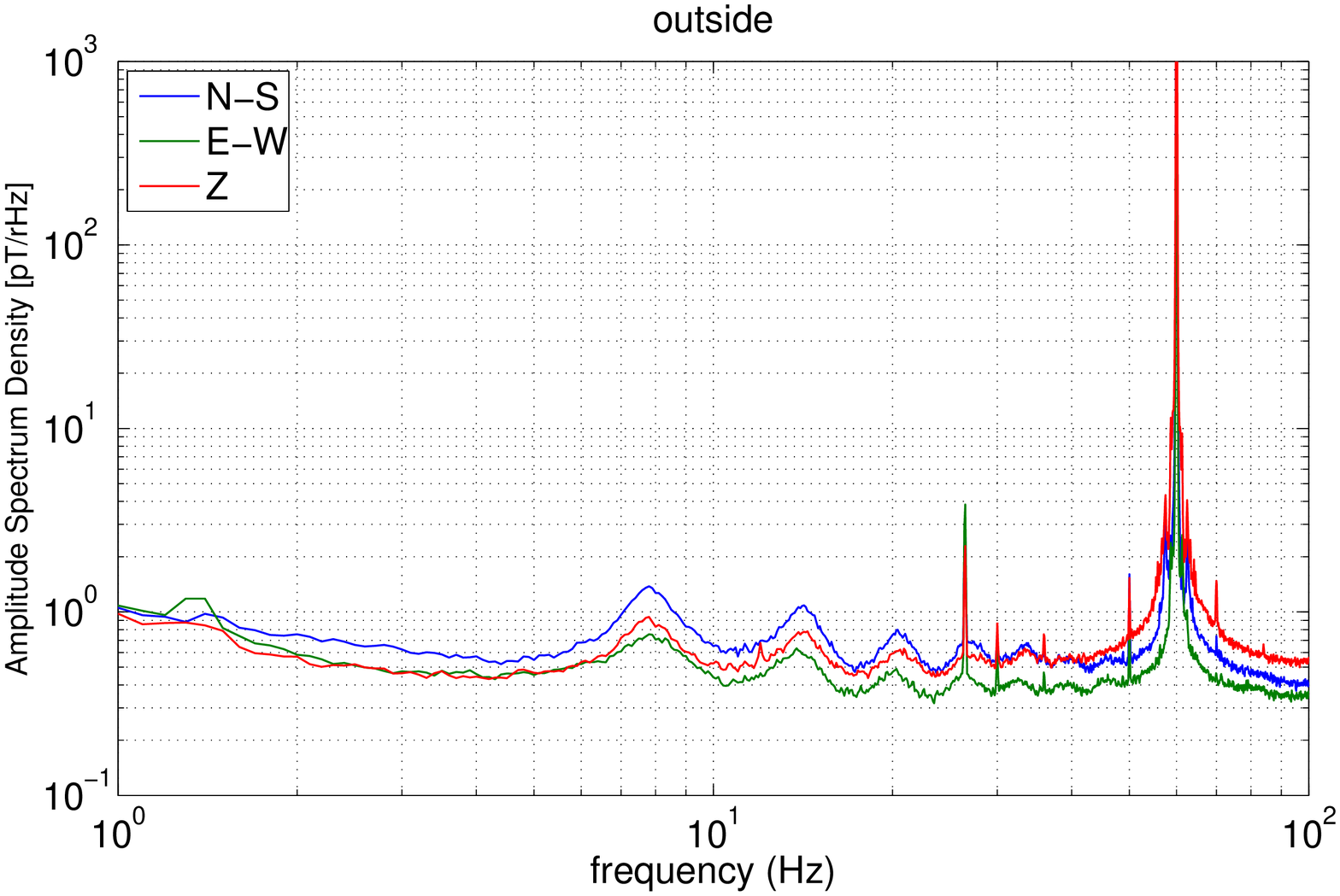}
\vspace{-0.2cm}
\caption{Spectrum of the magnetic field. The left and right panel show the spectrum measured inside and outside the KAGRA mine, respectively.}
\label{fig:mag_out}
\end{center}
\end{figure}

Figure \ref{fig:mag_in} shows the ratio of the spectrum of the magnetic fields inside and outside of the mine. One can see that the magnetic field inside the mine is larger than that outside the mine across a broad frequency band. In regard to the significant amplification of the magnitude along the Z direction, we can not deny the possibility that the amplification comes 
from the effect of artifacts such as electric instruments or metallic 
facilities around the center room. In order to understand this 
phenomenon, it will be required to make careful measurements of the magnetic 
fields at several locations in the mine, and to build and confirm a model 
to find a reason for the amplification.

\begin{figure}
\begin{center}
\includegraphics[width=0.9\linewidth]{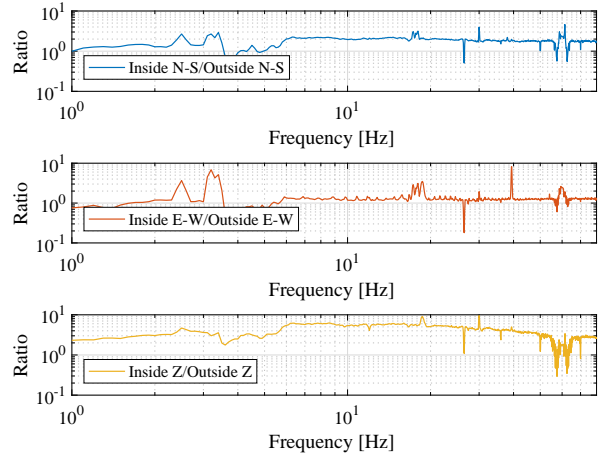}
\vspace{-0.2cm}
\caption{Ratio of the magnetic field spectra inside and outside the KAGRA mine.}
\label{fig:mag_in}
\end{center}
\end{figure}

\section*{Acknowledgement}
This work was supported by MEXT, JSPS Leading-edge Research Infrastructure Program, JSPS Grant-in-Aid for Specially Promoted Research 26000005, MEXT Grant-in-Aid for Scientific Research on Innovative Areas 24103005, JSPS Core-to-Core Program, A. Advanced Research Networks, and the joint research program of the Institute for Cosmic Ray Research. The magnetic field measurement shown in App.~\ref{sec:app2} was supported by the Joint Usage/ Research Center program of Earthquake Research Institute, the University of Tokyo. A part of the work was supported by the Advanced Technology Center (ATC) in the National Astronomical Observatory of Japan. A part of the work was supported by National Research Foundation (NRF) and Computing Infrastructure Project of KISTI-GSDC in Korea. This research activity has been also supported by the European Commission within the Seventh Framework Programme (FP7) - Project ELiTES (GA 295153). The authors would like to express our sincere appreciation to our colleagues in the LIGO and Virgo groups for their varied and continuing support. The authors would like to thank Albrecht R\"{u}diger for editorial support of this paper. Additionally, the authors also appreciate the intense efforts of the secretaries in the ICRR GW project office.


\begin{thebibliography}{10}
\bibitem{KAGRAname} KAGRA was originally called LCGT (Large-scale cryogenic gravitational wave telescope). After official approval of the project, we have added a nickname of KAGRA. KAGRA is named after the Japanese word for traditional sacred music and dance for the gods.
\end{thebibliography}
\end{document}